\documentclass[conference]{IEEEtran}
\IEEEoverridecommandlockouts
\usepackage{cite}
\usepackage{amsmath,amssymb,amsfonts}
\usepackage{algorithmic}
\usepackage{graphicx}
\usepackage{textcomp}
\usepackage{xcolor}
\usepackage{amsthm}
\usepackage{subcaption}
\usepackage{balance}

\theoremstyle{plain}
\newtheorem{theorem}{Theorem}

\newtheorem{corollary}[theorem]{Corollary}

\theoremstyle{definition}
\newtheorem{definition}[theorem]{Definition}

\theoremstyle{remark}

\def\BibTeX{{\rm B\kern-.05em{\sc i\kern-.025em b}\kern-.08em
    T\kern-.1667em\lower.7ex\hbox{E}\kern-.125emX}}
\begin{document}

\title{An Information-Theoretic Analysis of Temporal GNNs
}

\author{\IEEEauthorblockN{1\textsuperscript{st} Amirmohammad Farzaneh}
\IEEEauthorblockA{\textit{Department of Engineering Science} \\
\textit{University of Oxford}\\
Oxford, UK \\
amirmohammad.farzaneh@eng.ox.ac.uk}
}

\maketitle

\begin{abstract}
Temporal Graph Neural Networks, a new and trending area of machine learning, suffers from a lack of formal analysis. In this paper, information theory is used as the primary tool to provide a framework for the analysis of temporal GNNs. For this reason, the concept of information bottleneck is used and adjusted to be suitable for a temporal analysis of such networks. To this end, a new definition for Mutual Information Rate is provided, and the potential use of this new metric in the analysis of temporal GNNs is studied.
\end{abstract}


\section{Introduction}

The study of Graph Neural Networks (GNNs) is a rapidly growing area of machine learning and artificial intelligence. First introduced by \cite{scarselli2008graph} back in 2009, it has now become one of the trendiest areas in machine learning research. The importance of this field becomes more evident by observing the amount of graphical data structures around us. Social networks, communication networks, and biological networks are only a few examples of such graphs. Graphs can be used to represent the relationship present between data in many different fields of science, and therefore GNNs have been used to solve problems in areas such as modern recommender systems, computer vision, natural language processing (NLP), software mining, bioinformatics, and urban intelligence \cite{GNNBook2022}.

Additionally, most of these graphs are dynamic by nature. This means that we expect them to change and evolve over time. For instance, new nodes and links are always being formed or deleted in social media networks. Because of this reason, Temporal GNNs have recently attracted a lot of attention from the machine learning community. Temporal GNNs are used to study and learn from graphical data and their evolution through time, whereas standard GNNs learn from single graphs. A survey on existing methods on this topic has been recently published \cite{kazemi2020representation}, and even newer methods have been introduced since then \cite{rossi2020temporal}.

Generally speaking, there is a lack of theoretical formulations to assess the performance of GNNs. Consequently, the temporal version of GNNs also suffers from the same lack of theoretical assessment in a more serious manner. Because of this, we propose an information-theoretic approach to analysing the performance of temporal graph learning algorithms. The information bottleneck principle is used as the core idea, while the temporal graph evolutions are modeled using stochastic processes. Mutual Information Rate (MIR) is then used as the main metric for analysing the performance of algorithms that work on temporal graphs. Additionally, a new and more robust metric to replace MIR is introduced, and it is discussed how NECs can be used for analysing temporal GNNs.

\section{The information bottleneck problem for temporal GNNs}

\begin{figure}[h!]
    \centering
    \includegraphics[width = 0.8\columnwidth]{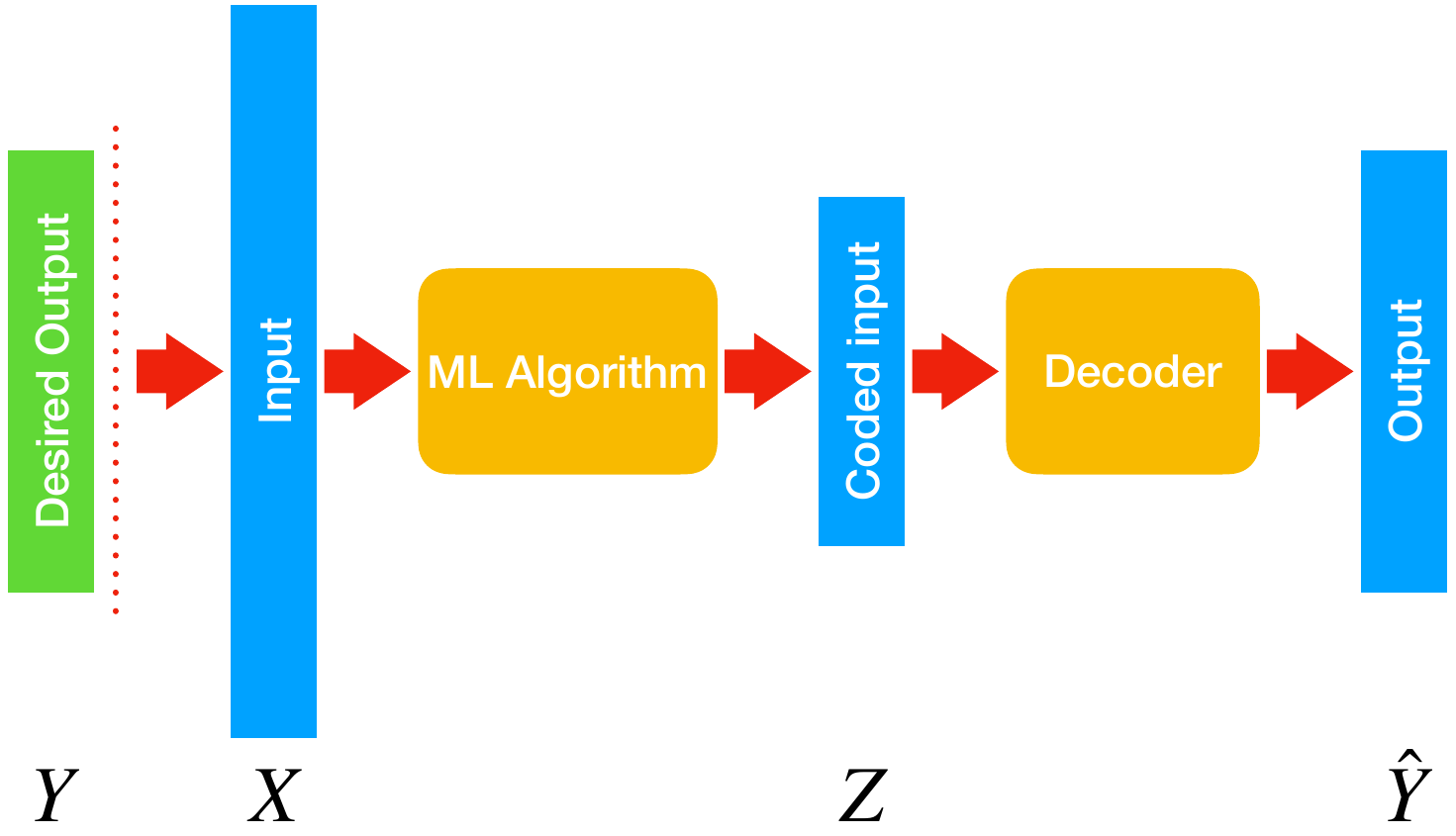}
    \caption{Model of an ML algorithm}
    \label{fig::ML_alg}
\end{figure}

The information-bottleneck problem \cite{goldfeld2020information} is often used as the information-theoretic formulation of Machine Learning. This method models the machine learning algorithm using an encoder-decoder pair, and formulates the problem as a trade-off between the compression ratio of the input and the accuracy of the output. We show the input to the algorithm with $X$, the desired output with $Y$, and the actual output of the algorithm with $\hat{Y}$. Additionally, we use $Z$ to show the output of the encoder. This will be the output of the last layer if we are working with a neural network \cite{shwartz2017opening}. Fig. \ref{fig::ML_alg} shows these notations in the context of a learning algorithm.

Using the model of Fig. \ref{fig::ML_alg}, the information bottleneck problem can be used to model the ML problem as below.

\begin{equation}
\begin{split}
    \textrm{inf}& \quad I(X;Z)\\
    \textrm{subject to}& \quad I(Y;Z)\geq\alpha,
\end{split}
\label{eqn::inf_bottleneck}
\end{equation}
where $I$ is used to show mutual information as per Shannon's definition \cite[Ch.~2]{cover1999elements}, and $\alpha$ is the parameter that sets the accuracy of the algorithm.

However, Eq. \ref{eqn::inf_bottleneck} can not be used in the context of dynamic graphs. This is because when dealing with temporal graphs, we should look at sequences of the input and output, rather than single ones. In this case, rather than assuming the variables in Eq. \ref{eqn::inf_bottleneck} to be single graphs, we will consider them to be stochastic processes. In other words we have
\begin{equation}
    \begin{split}
        Y &= \{Y_1, Y_2, \ldots, Y_n\},\\
        X &= \{X_1, X_2, \ldots, X_n\},\\
        Z &= \{Z_1, Z_2, \ldots, Z_n\},\\
        \hat{Y} &= \{\hat{Y}_1, \hat{Y}_2, \ldots, \hat{Y}_n\},\\
    \end{split}
\end{equation}
where the following Markov chain governs each $i$, which is a time snapshot of the network.
\begin{equation}
    Y_i\rightarrow X_i \rightarrow Z_i \rightarrow \hat{Y}_i
\end{equation}

Eq. \ref{eqn::inf_bottleneck} tells us that in order to assess an ML algorithm we can look at the mutual information between $X$ and $Z$, and $Y$ and $Z$, respectively. However, in a dynamic graph setup, we have $n$ time snapshots of these random variables. The natural question that arises in these scenario is the value of $i$ to look at. In this case, we will use the concept of MIR \cite{gray1980mutual}.

When working with stochastic processes, we can look at the mutual information between pairs of corresponding variables, or $I(X_1,\ldots,X_n;Y_1,\ldots,Y_n)$. However, the value of this can theoretically be unlimited as the process grows large. Therefore, we rely on the definition of MIR, which measures average mutual information per unit of time defined as below.

\begin{definition}[Mutual Information Rate (MIR)]
\label{def::MIR}
The mutual information rate between two stochastic processes $X$ and $Y$ is defined as below.
\[
\textrm{MIR} = \lim_{n \to \infty}\frac{1}{n}I\left(X_1,X_2,\ldots, X_n;Y_1, Y_2, \ldots, Y_n\right)
\]
\end{definition}

Having defined the variables in Fig. \ref{fig::ML_alg} and using the notions from the information bottleneck principle, we propose the following metrics to be used for the assessment of temporal graph learning algorithms.
\begin{itemize}
    \item \textbf{Accuracy:} $\textrm{MIR}(Y;Z)$
    \item \textbf{Compression:} $\textrm{MIR}(X;Z)$
    \item \textbf{Overall score:} $\textrm{MIR}(Y;Z)/\textrm{MIR}(X;Z)$
\end{itemize}

Note that there exist numerous methods for estimating the MIR of a pair of stochastic processes. For instance, the mutual information estimation setup proposed in \cite{goldfeld2020information}[p.~17] can be used. However, because of the existence of $2n$ random variables in the definition of MIR, especially as $n$ grows large, we propose an alternative metric for measuring the rate of mutual information.

In the remainder of this section, we will use the following notation for better readability when referring to subsets of a stochastic process.

\begin{equation}
    X_{i:j} \equiv \{X_i,X_{i+1}, \ldots, X_j\}
\end{equation}

\section{Alternative Mutual Information Rate (AMIR)}

The current definition for MIR makes its applications to stochastic processes very limited, as it involves the mutual information of the entire process as it grows large. This notion of MIR might not be feasible to study for real-world processes, as estimating the mutual information of such big stochastic processes might not be accurate. Because of this, we propose the definition of a novel relative metric for the mutual information rate between processes. This new metric is called alternative mutual information rate, or AMIR for short. Even though a similar metric exists for entropy rate of stochastic processes, the absence of this metric for analysing mutual information can be felt. This novel metric analyses the mutual information of single random variables, given the history of the sequence. We believe that this novel metric can be estimated more accurately, especially if the given stochastic processes are Markov chains.

Firstly, note that the definition of MIR as stated in Definition \ref{def::MIR} is very similar to that of entropy rate \cite[Ch.~4]{cover1999elements}. Additionally, there exists a relative version of entropy rate, which is shown to be equal to entropy rate for stationary processes \cite[Thm.~4.2.1]{cover1999elements}. This new relative version of the entropy rate had the benefit of being easier to work with and use in our estimations and proofs. Based on the similarities between the concepts of entropy rate and MIR, we search for a relative format for MIR as well. This conditional concept of MIR can help us in many applications that involve the correlation of two stochastic processes, where we want to study the limit of the behaviour of single random variables in a conditional manner. Based on this, we provide the following definition for this alternative version of MIR.

\begin{definition}[Alternative Mutual Information Rate (AMIR)]
 The Alternative Mutual Information Rate (AMIR) between two stochastic processes $X$ and $Y$ is defined as
 \begin{equation}
     \text{AMIR}(X;Y) = \lim_{n\to \infty} I(X_n; Y_n|X_{1:n-1}, Y_{1:n-1}),
 \end{equation}
 when it exists.
\end{definition}

In the remainder of this section, we will look at the properties of AMIR, its existence, and its relationship with MIR. We first start with the following theorem.

\begin{theorem}
\label{limit}
For two stationary processes $X$ and $Y$, AMIR exists.
\end{theorem}
\begin{proof}
    It is well known that for two stationary processes $X$ and $Y$, $\text{MIR}(X;Y)$ exists \cite{gray1980mutual}. For AMIR, we can write
    \begin{equation}
    \label{AMIR1}
        \begin{split}
            \text{AMIR}(X;Y) &= \lim_{n\to \infty}I(X_n;Y_n|X_{1:n-1},Y_{1:n-1})\\
            & = \lim_{n\to \infty} H(X_n|X_{1:n-1},Y_{1:n-1}) \\
            &+ H(Y_n|X_{1:n-1},Y_{1:n-1}) \\
            &- H(X_n,Y_n|X_{1:n-1},Y_{1:n-1}).\\
        \end{split}
    \end{equation}

Firstly, note that based on the stationarity of $X$, $Y$, and $(X,Y)$, we have the following equation \cite[Ch.~4]{cover1999elements}.

\begin{equation}
\label{AMIRXY}
    \lim_{n \to \infty} H(X_n,Y_n|X_{1:n-1},Y_{1:n-1}) = \lim_{n \to \infty}\frac{1}{n}H(X_{1:n},Y_{1:n})
\end{equation}

Eq. \ref{AMIRXY} shows that $H(X_n,Y_n|X_{1:n-1},Y_{1:n-1})$ tends towards the entropy rate of $(X,Y)$, and has a limit. Therefore, in order to show that the limit for AMIR exists, we only need to show that the other two limits in Eq. \ref{AMIR1} exist. For $H(X_n|X_{1:n-1},Y_{1:n-1})$ we have
\begin{equation}
    \begin{split}
        H(X_n|X_{1:n-1},Y_{1:n-1})&\leq H(X_n|X_{2:n-1},Y_{2:n-1})\\
        & = H(X_{n-1}|X_{1:n-2},Y_{1:n-2}),
    \end{split}
\end{equation}
where the first line comes from the fact that conditioning reduces entropy, and the second line can be concluded from the stationarity of both processes. Therefore, $H(X_n|X_{1:n-1},Y_{1:n-1})$ forms a non-increasing sequence of positive numbers, which has a limit. A similar approach can be used to prove that $H(Y_n|X_{1:n-1},Y_{1:n-1})$ also converges. Therefore, AMIR also has a limit and is well-defined for stationary processes.
\end{proof}

Additionally, we can state the following corollary for Theorem \ref{limit}.

\begin{corollary}
\label{lessthan}
For stationary processes $X$ and $Y$ we always have
\begin{equation}
    \text{AMIR}(X;Y) \leq \text{MIR}(X;Y).
\end{equation}
\end{corollary}
\begin{proof}
    We can use Eq. \ref{AMIR1} and write the following chain of equations.
    \begin{align}
    \label{AMIR_less}
        \begin{split}
            \text{AMIR}(X;Y) & = \lim_{n\to \infty} H(X_n|X_{1:n-1},Y_{1:n-1}) \\
            &+ H(Y_n|X_{1:n-1},Y_{1:n-1}) \\
            &- H(X_n,Y_n|X_{1:n-1},Y_{1:n-1})
        \end{split}
        \\
        \begin{split}
            & = \lim_{n\to \infty} H(X_n|X_{1:n-1}) \\
            &- I(X_n;Y_{1:n-1}) \\
            &+ H(Y_n|Y_{1:n-1};X_{1:n-1}) \\
            &- I(Y_n;X_{1:n-1})\\
       &- H(X_n,Y_n|X_{1:n-1},Y_{1:n-1})
        \end{split}
        \\
        \begin{split}
        & = H(X)+H(Y)-H(X,Y)\\
        &-I(X_n;Y_{1:n-1})- I(Y_n;X_{1:n-1})
        \end{split}
        \\
        \label{AMIR_less2}
        \begin{split}
        & = \text{MIR}(X;Y)\\
        &-I(X_n;Y_{1:n-1})- I(Y_n;X_{1:n-1})
        \end{split}
    \end{align}
    Based on the fact that mutual information is non-negative, we can easily conclude the claim of the corollary based on Eq. \ref{AMIR_less2}.
\end{proof}

Additionally, the conditions under which MIR and AMIR are equal are stated in the following corollary.

\begin{corollary}
\label{equality}
    For stationary processes $X$ and $Y$, MIR and AMIR are equal if and only if the following conditions are met.
    \begin{enumerate}
        \item Given $X_{1:n-1}$, $X_n$ and $Y_{1:n-1}$ are independent.
        \item Given $Y_{1:n-1}$, $Y_n$ and $X_{1:n-1}$ are independent.
    \end{enumerate}
\end{corollary}
\begin{proof}
    As $I(X_n;Y_{1:n-1}|X_{1:n-1})$ and $I(Y_n;X_{1:n-1}|Y_{1:n-1})$ are non-negative, the only way that the values of MIR and AMIR can be equal according to Eq. \ref{AMIR_less2} is if these two values are zero. For $I(X_n;Y_{1:n-1}|X_{1:n-1})$ we can write
    \begin{equation}
        \begin{split}
            &I(X_n;Y_{1:n-1}|X_{1:n-1}) = 0\\
            &\Rightarrow H(X_n|X_{1:n-1})-H(X_n|X_{1:n-1},Y_{1:n-1}) = 0\\
            &\Rightarrow H(X_n|X_{1:n-1})=H(X_n|X_{1:n-1},Y_{1:n-1}),
        \end{split}
    \end{equation}
    which translates into $X_n$ and $Y_{1:n-1}$ being independent given $X_{1:n-1}$. The other condition can be obtained by setting $I(Y_n;X_{n-1}|Y_{n-1})$ to zero.
\end{proof}

It can be seen that in addition to the processes being stationary, the conditions in Corollary \ref{equality} need to be met in order for MIR and AMIR to be equal. This is unlike the equality of entropy rate and relative entropy rate, where the only condition was the stationarity of the process. However, it can also be seen that the conditions for the equality of these two metrics are not very strict, and can be assumed for many real-life pairs of stochastic processes.

\section{Benefits of AMIR}
\label{benefits}
In this section, we will briefly mention and discuss two advantages of AMIR over MIR.

\subsection{Rate of convergence}

Existing methods for estimating mutual information of random variables rely on samples from the distribution of those random variables \cite{kraskov2004estimating, carrara2020estimation}. Therefore, as the space of possible data points grows, more sample are needed to accurately estimate this metric. Looking at the definition of MIR and AMIR, it can be seen that the space of both metrics includes $(|X||Y|)^n$ possibilities. However, there are usually assumptions in place about the memory of the system, which dictates how many samples in the future will be affected by the current sample. Based on this, there usually exists a constant C for which we can assume
\begin{multline}
    I(X_n;Y_n|X_{1:n-1}, Y_{1:n-1}) \\
    = I(X_n;Y_n|X_{n-C:n-1}, Y_{n-C:n-1}).
\end{multline}
This assumption will reduce the size of the sample space to $(|X||Y|)^{C+1}$, which lowers the number of samples needed to estimate AMIR and increase the convergence rate.

\subsection{Number of sample sequences needed}
It must be noted that in order to estimate MIR, we need samples of $(X_n,Y_n)$, which means samples of both sequences. Additionally, the number of required samples for accurate estimation will increase exponentially with $n$. However, with AMIR, one sequence for $(X,Y)$ will suffice if the memory of the system is assumed to be limited. In other words, with every new sample for the sequence, we will have a new sample for estimating $I(X_n,Y_n|X_{n-C:n-1}, Y_{n-C:n-1})$, which removes the need for having numerous sequences.

\section{AMIR of Markov chains}

In this section, we will consider the Mutual Information Rate of stationary Markov chains. As a special case of stochastic processes, Markov chains have applications in many different areas.

Consider $X$ and $Y$ to be stationary Markov chains. Let $\mu_X$, $\mu_Y$, and $\mu_{XY}$ be the stationary distributions of $X$, $Y$, and $(X,Y)$, respectively. Additionally, let $P$, $Q$, and $R$ be the transition probability matrices for $X$, $Y$, and $(X,Y)$, respectively. To calculate the MIR of $X$ and $Y$, we can write
\begin{equation}
\label{MIR_Markov}
\begin{split}
    \text{MIR}(X;Y) &= H(X)+H(Y)-H(X,Y)\\
    & = -\sum_{ij}\mu_{X,i}P_{ij}\log P_{ij} \\
    &-\sum_{ij}\mu_{Y,i}Q_{ij}\log Q_{ij}\\
    &+\sum_{ij}\mu_{XY,i}R_{ij}\log R_{ij}.
\end{split}
\end{equation}
Note that the entropy rate of stationary Markov chains can be found using \cite[Thm.~4.2.4]{cover1999elements}.

Additionally, it can be seen that the condition of causality holds for most of the processes that are simulated using Markov chains. For instance, Markov chains have recently been used for temporal alignment of medical signals in dementia \cite{COSTA2023104328}. In this study, it is safe to assume that future signals will not affect the past, and therefore the assumption of causality holds. Consequently, it is safe to state that for most of the Markov chains we work with, MIR and AMIR will be equal, but it is always necessary to check the condition of Corollary \ref{equality}.

Even though both MIR and AMIR can be calculated using Eq. \ref{MIR_Markov}, we do not necessarily always have all the parameters of the Markov chain, and therefore can not use Eq. \ref{MIR_Markov} directly. Therefore, we need to rely on estimation. We believe that in estimation, AMIR can prove to be a more robust and reliable estimator compared to MIR, as much less parameters are involved. Observe that for MIR, we need to estimate $I(X_1,\ldots,X_n;Y_1,\ldots, Y_n)$, whereas the value to estimate for AMIR in Markov chains is simply $I(X_n;Y_n|X_{n-1},Y_{n-1})$.

\section{Simulation results}
\label{simulations}
In this section, we will perform simulations on a number of stochastic processes to compare the performance of MIR and AMIR as metrics of mutual information to study stochastic processes.

\subsection{Gaussian processes}
In this section, we simulate MIR and AMIR for two correlated Gaussian processes. For this purpose, we chose the following processes.
\begin{itemize}
    \item $X_i\sim \mathcal{N}(0,\,1)$
    \item $Y_i\sim 0.8X_i + 0.6\mathcal{N}(0,\,1)$
\end{itemize}
It can be checked that as the variables in these processes are i.i.d, they satisfy the conditions of Corollary \ref{equality}, and MIR and AMIR are equal for these two processes. Additionally, both MIR and AMIR are simply equal to the mutual information of a pair of these random variables, which can be calculated as below.
\begin{equation}
\begin{split}
    I(X_i,Y_i) &= H(Y_i)-H(Y_i|X_i)\\
    & = H(\mathcal{N}(0,\,1)) - H(\mathcal{N}(0,\,0.36))\\
    & = 0.5\log \frac{1}{0.36}
\end{split}
\end{equation}

\begin{figure}
\vspace{1pt}
\centering     
    \begin{subfigure}{\columnwidth}
         \centering
         \includegraphics[width=0.8\textwidth]{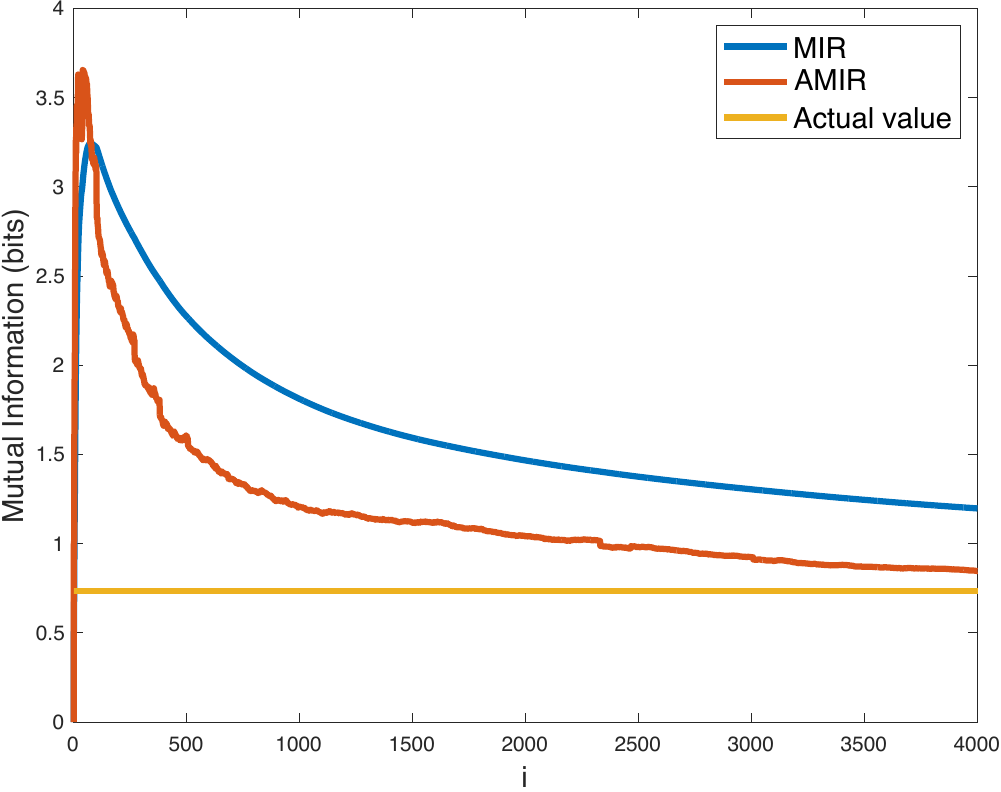}
         \caption{The estimations for MIR and AMIR}
         \label{sim1a}
     \end{subfigure}
     \hfill
    \begin{subfigure}{\columnwidth}
         \centering
         \includegraphics[width=0.8\textwidth]{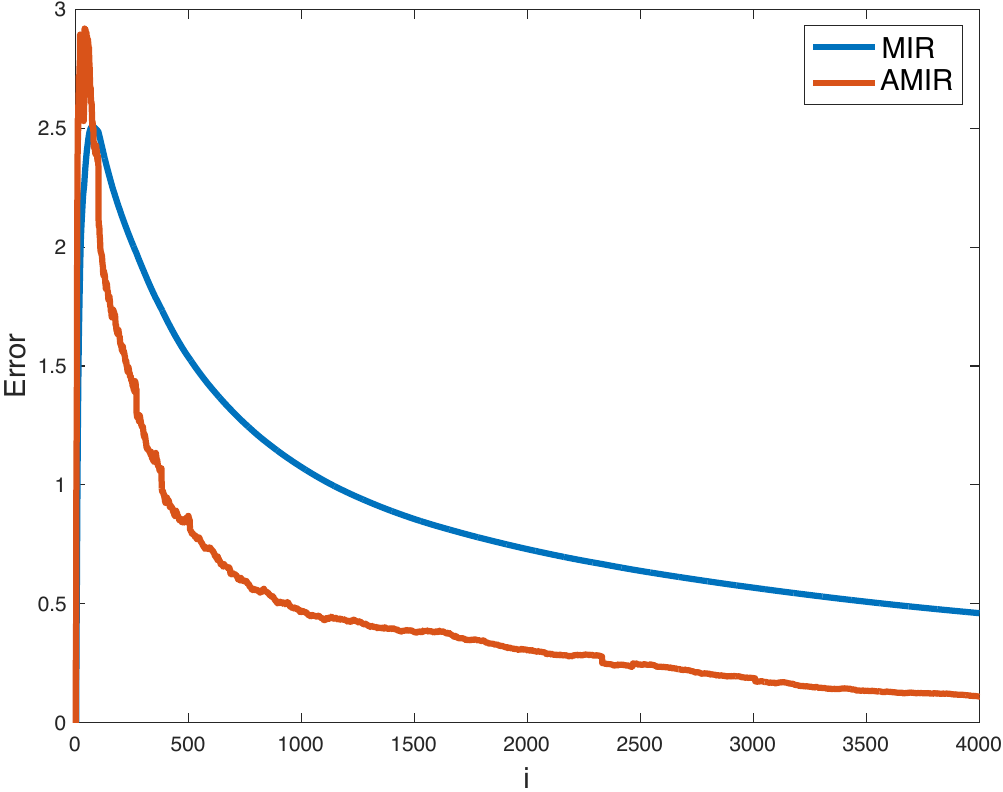}
         \caption{Error in the convergence of MIR and AMIR}
         \label{sim1b}
     \end{subfigure}
     \hfill
\caption{Estimating MIR and AMIR}
\label{simulation1}
\end{figure}

Fig. \ref{simulation1} shows the result of this simulation for estimating MIR and AMIR. Fig. \ref{sim1a} shows the estimate for MIR and AMIR as the sequence grows large, alongside the actual calculated value for the mutual information. Fig. \ref{sim1b} shows the error in the estimate as the difference between the estimated values for MIR and AMIR with the actual value. It can be seen that even though AMIR shows more fluctuations, it is faster in converging to the theoretical limit. This is in accordance with our predictions about the behaviour of AMIR.

\subsection{Hidden Markov Model}

In this section, we created and simulated a two state Hidden Markov Model (HMM) $(X,Y)$ with the following properties.
\begin{equation}
\label{HMM_params}
    \begin{cases}
    \mathbf{Z}_X = \begin{bmatrix}
        0.8 & 0.2\\
        0.4&0.6
    \end{bmatrix}\\
    p(Y_i|X_i) = \begin{bmatrix}
        0.7 & 0.3\\
        0.3&0.7
    \end{bmatrix}
    \end{cases}
\end{equation}

Firstly, it must be noted that this pair of stationary processes does not satisfy the conditions of Corollary \ref{equality}. This is because it can be checked that $Y_i$ is not independent from $X_{i-1}$. Therefore, we expect MIR and AMIR to be different for this pair of processes. Based on the parameters given in Eq. \ref{HMM_params}, the values for MIR and AMIR can be calculated, and have the following values.

\begin{equation}
    \begin{cases}
        \text{MIR}(X;Y) = 0.1035 \text{ bits}\\
        \text{AMIR}(X;Y) = 0.0892 \text{ bits}\\
    \end{cases}
\end{equation}

\begin{figure}
\vspace{1pt}
    \centering
    \includegraphics[width=0.8\columnwidth]{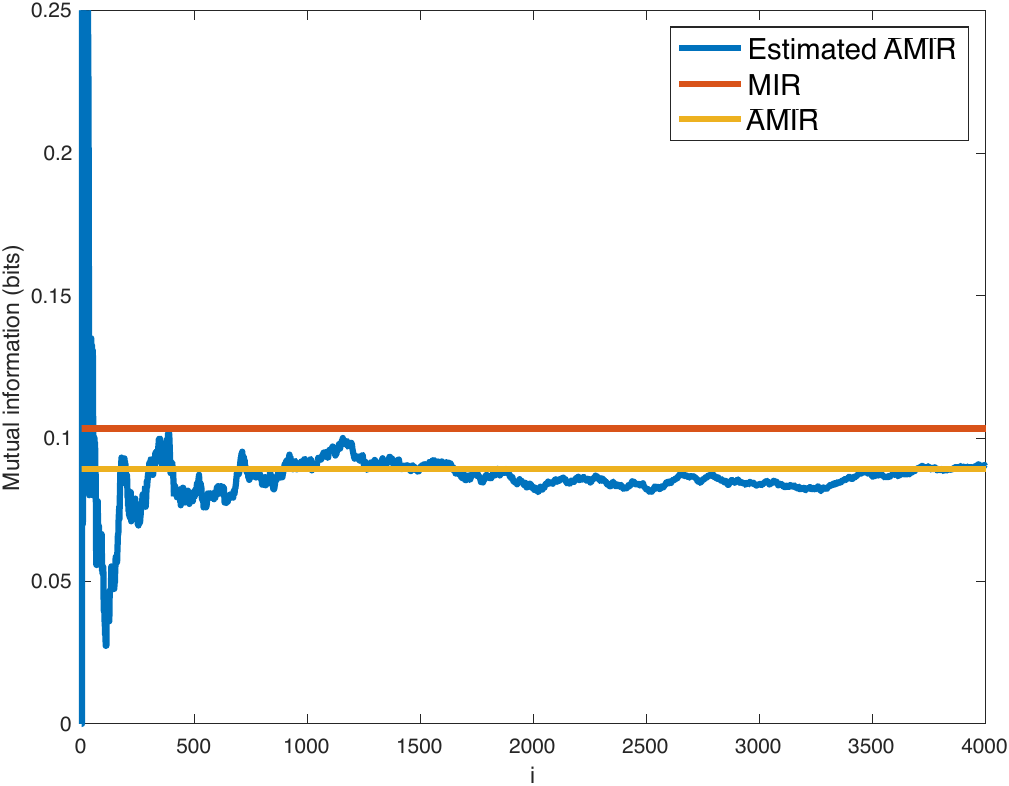}
    \caption{Estimating AMIR of an HMM}
    \label{HMM_sim}
\end{figure}

We then ran the simulation for one random instance of the HMM with a length of 4000, in order to estimate the value for AMIR using its definition. The results of this simulation are illustrated in Fig. \ref{HMM_sim}. The first thing that can be observed is the fact that the estimated AMIR is clearly tending towards its theoretical limit, as proven in Theorem \ref{limit}. Secondly, it can be seen that as stated in Corollary \ref{lessthan}, the value for AMIR is less than MIR. Ultimately, this result is showcasing one of the main advantages of AMIR, which is that it can be estimated from having a single sequence from both processes. However, to estimate MIR from observed data, we are in need of numerous instances of both sequences.  

\section{Application of Network Evolution Chains to temporal GNNs}

It can be seen that due to the similarities between the concepts of MIR and entropy rate, Network Evolution Chains \cite{farzaneh2022information} can be used to simulate the sequence of input graphs, and the information-theoretic analysis done in \cite{farzaneh2022information} can be employed to study temporal GNNs. Using Network Evolution Chains to model the input sequence of graphs will impose a Markov chain assumption on the input, which will result in the following simplification for MIR.

\begin{equation}
    \textrm{MIR} = \lim_{n \to \infty}I\left(X_n;Y_n|X_{n-1},Y_{n-1}\right)
\end{equation}

Additionally, there are parallels to all the concepts discussed in \cite{farzaneh2022information} in terms of MIR. For instance, an equivalent of the AEP can be stated for MIR in the following format.

\begin{equation}
\label{AEP for MIR}
    \frac{1}{n}\log \frac{p(X_{1:n}, Y_{1:n})}{p(X_{1:n})p(Y_{1:n})} \rightarrow \text{MIR}(X;Y) \quad \text{with probability 1}
\end{equation}

The proof for Eq. \ref{AEP for MIR} can be done in a similar way to the original AEP, which can be found in \cite[Thm.~16.8.1]{cover1999elements}.   

If the input graphs to a temporal GNN are modeled using an NEC, then it satisfies the conditions for the AEP for MIR. Additionally, as the relative version of MIR, AMIR can be used to calculate the MIR of the process, just like the approach done in \cite{farzaneh2022information}. This then gives rise to the extension of all the concepts discussed in \cite{farzaneh2022information} in the context of temporal GNNs and MIR. This goes beyond the scope of this paper, and is left as a future work on this subject.

\section{Conclusion}

In this paper, we provided the information-theoretic fundamentals of the analysis of temporal GNNs. After emphasizing on the importance and relevance of GNNs, we argued why there has recently been a lot of research on the temporal analysis of such networks. Because of the lack of a theoretical framework for analysing temporal GNNs, we introduced an information-theoretic methodology to assess the performance of temporal GNNs. Finally, we argued why simulating the temporal evolution of graphs using NECs can further improve the benefits of the introduced model, and expand its applications.

\section*{Acknowledgment}

For the purpose of open access, the author has applied a creative commons attribution (CC BY) licence (where permitted by UKRI, ‘open government licence’ or ‘creative commons attribution no-derivatives (CC BY-ND) licence’ may be stated instead) to any author accepted manuscript version arising.

I would like to thank Professor Justin Coon and Dr. Mihai-Alin Badiu for their invaluable help and support with this project.

\balance

\end{document}